\documentclass[prb,twocolumn,showpacs,amsmath,amssymb,superscriptaddress,bibnotes,floatfix]{revtex4-2}
\usepackage[dvipsnames]{xcolor}
\usepackage[utf8]{inputenc}
\usepackage{graphicx}
\usepackage[english]{babel}
\usepackage{amsmath}
\usepackage{amssymb}
\usepackage{amsthm}
\usepackage{braket}
\usepackage{tikz}
\usepackage{adjustbox}
\usepackage{bbold}
\usepackage{algorithm}
\usepackage{algpseudocode}
\usepackage[justification=raggedright,singlelinecheck=true]{caption}

\usepackage{subcaption}
\usepackage[percent]{overpic} 
\usepackage{nth}
\usepackage{calc}
\usepackage{hyperref}
\usepackage{cleveref}
\usepackage{placeins}

\crefname{equation}{Eq.}{Eqs.}

\graphicspath{{thesis/diagrams/L4_evo/}{thesis/diagrams/}{thesis/figures_used}{figures}}

\newcommand{\defaulttensorsize}{10pt}
\newcommand{\tensorsize}{\defaulttensorsize}

\usetikzlibrary{patterns,patterns.meta}
\usetikzlibrary{shapes.geometric}
\usetikzlibrary{shapes.misc}
\tikzstyle{tensor}=[draw, inner sep=0, outer sep=0, minimum size=\tensorsize]
\tikzstyle{notensor}=[inner sep=0, outer xsep=2pt, outer ysep=0, minimum size=\tensorsize]
\tikzstyle{atensor}=[tensor, circle]
\tikzstyle{ctensor}=[tensor, circle]
\tikzstyle{dtensor}=[tensor, diamond]
\tikzstyle{wtensor}=[tensor]
\tikzstyle{etensor}=[tensor, minimum height=(1cm/\defaulttensorsize*0.5*2+1)*\tensorsize]
\tikzstyle{widetensor}[2]=[tensor, minimum width=(1cm/\defaulttensorsize*0.75*(#1-1)+1)*\tensorsize]
\tikzstyle{leg}=[draw, thick]
\tikzstyle{idx_label}=[font=\scriptsize]
\newcommand{\stripesize}{4pt}
\pgfdeclarepatternformonly{stripes}{\pgfpointorigin}{\pgfpoint{\stripesize}{\stripesize}}{\pgfpoint{\stripesize}{\stripesize}}
{
    \pgfpathmoveto{\pgfpoint{0}{0}}
    \pgfpathlineto{\pgfpoint{\stripesize}{\stripesize}}
    \pgfpathlineto{\pgfpoint{\stripesize}{0.5*\stripesize}}
    \pgfpathlineto{\pgfpoint{0.5*\stripesize}{0}}
    \pgfpathclose%
    \pgfusepath{fill}
    \pgfpathmoveto{\pgfpoint{0}{0.5*\stripesize}}
    \pgfpathlineto{\pgfpoint{0}{\stripesize}}
    \pgfpathlineto{\pgfpoint{0.5*\stripesize}{\stripesize}}
    \pgfpathclose%
    \pgfusepath{fill}
}
\tikzstyle{striped}=[pattern=stripes, pattern color=lightgray]

\tikzstyle{tensornetwork}=[baseline=-0.25em, xscale=0.75, yscale=0.5,
                           every node/.style={inner sep=0, outer xsep=2pt, outer ysep=5pt, text depth=0pt},
                           execute at end picture={\path (current bounding box.south west) +(-2pt, 0) (current bounding box.north east) +(2pt, 0);}]

\newcommand{\vimg}[2][0.5]{%
    \adjustbox{valign=c}{\includegraphics[scale=#1]{#2}}%
}

\newcommand{\tikzTreeOpt}{
		\begin{tikzpicture}[tensornetwork]
			\node[atensor]    (A1) at (0,-1) {};
			\node[atensor]    (A2) at (0,1) {};
			\draw (A1) -- (A2);

			\draw (A1.west) -- +(-0.5, 0) node[left, idx_label] {$k$};
			\draw (A2.west) -- +(-0.5, 0) node[left, idx_label] {$i$};
			\draw (A1.east) -- +(0.5, 0) node[right, idx_label] {$l$};
			\draw (A2.east) -- +(0.5, 0) node[right, idx_label] {$j$};
		\end{tikzpicture}
		=
		\begin{tikzpicture}[tensornetwork]
			\node[etensor]    (A1) at (0,0) {$\phi$};
			\node[notensor]    (A2) at (-1,1) {};
			\node[notensor]    (A3) at (-1,-1) {};
			\node[notensor]    (A4) at (1,1) {};
			\node[notensor]    (A5) at (1,-1) {};

			\draw  (A1.west |- A2) -- (A2) node[left, idx_label] {$i$};
			\draw  (A1.west |- A3) -- (A3) node[left, idx_label] {$k$};
			\draw  (A1.east |- A4) -- (A4) node[right, idx_label] {$j$};
			\draw  (A1.east |- A5) -- (A5) node[right, idx_label] {$l$};
		\end{tikzpicture}
}

\newcommand{\tikzTreeOptPoss}{
		\begin{tikzpicture}[tensornetwork, node distance=1cm]
			\node[atensor] (A1) at (0,2) {};
			\node[atensor] (A2) at (0,0) {};
			\node[dtensor, red] (D1) at (0,1) {};

			\draw (A1.west) -- +(-0.5,0) node[left, idx_label] {$i$};
			\draw (A1.east) -- +(0.5,0) node[right, idx_label] {$j$};
			\draw (A2.west) -- +(-0.5,0) node[left, idx_label] {$k$};
			\draw (A2.east) -- +(0.5,0) node[right, idx_label] {$l$};

			\draw (A1.south) -- (D1.north);
			\draw (A2.north) -- (D1.south);

			\begin{scope}[xshift=3.5cm]
				\node[atensor] (A1) at (0,2) {};
				\node[atensor] (A2) at (0,0) {};
				\node[dtensor, red] (D1) at (0,1) {};

				\draw (A1.west) -- +(-0.5,0) node[left, idx_label] {$i$};
				\draw (A1.east) -- +(0.5,-2) node[right, idx_label] {$l$};
				\draw (A2.west) -- +(-0.5,0) node[left, idx_label] {$k$};
				\draw (A2.east) -- +(0.5,2) node[right, idx_label] {$j$};

				\draw (A1.south) -- (D1.north);
				\draw (A2.north) -- (D1.south);
			\end{scope}
			\begin{scope}[xshift=6cm]
				\node[atensor] (A5) at (0,1) {};
				\node[atensor] (A6) at (2,1) {};
				\node[dtensor, red] (D3) at (1,1) {};

				\draw (A5.north) -- +(0,0.5) node[above, idx_label] {$i$};
				\draw (A5.south) -- +(0,-0.5) node[below, idx_label] {$k$};
				\draw (A6.north) -- +(0,0.5) node[above, idx_label] {$j$};
				\draw (A6.south) -- +(0,-0.5) node[below, idx_label] {$l$};

				\draw (A5.east) -- (D3.west);
				\draw (A6.west) -- (D3.east);
			\end{scope}
		\end{tikzpicture}
}

\begin{document}
\preprint{APS/123-QED}
\title{Tree tensor networks for many-body localization in two dimensions}
\author{Lars Humpert}%
 \affiliation{Institute for Theory of Statistical Physics, RWTH Aachen University and JARA—Fundamentals of Future Information
Technology, 52056, Aachen, Germany}
 
\author{Dante M.~Kennes}

 \affiliation{Institute for Theory of Statistical Physics, RWTH Aachen University and JARA—Fundamentals of Future Information
Technology, 52056, Aachen, Germany}
 \affiliation{Max Planck Institute for the Structure and Dynamics of Matter,
Center for Free Electron Laser Science, Luruper Chaussee 149, 22761 Hamburg, Germany}

\author{Jan-Niklas Herre}%
 \email{jan.herre@rwth-aachen.de}
 \affiliation{Institute for Theory of Statistical Physics, RWTH Aachen University and JARA—Fundamentals of Future Information
Technology, 52056, Aachen, Germany}
\date{\today}

\begin{abstract}
We investigate the disordered spin-$\frac12$Heisenberg model in two dimensions and employ tree tensor networks (TTNs) with a physics-informed structural optimization of the tree layout, to simulate dynamics in the many-body localization problem. We find that TTNs are able to capture two-dimensional entanglement patterns more effectively than matrix product states (MPS) while being more efficient to contract than projected entangled pair states (PEPS) to probe larger systems and longer times. Structural optimization of the trees based on time evolution of the entanglement in the system allows to keep the necessary bond dimensions low and to maximally exploit the increased expressiveness of TTNs over MPS. In this way, we achieve more accurate results in all considered parameter regimes both below and above the ergodicity-to-localization crossover at a comparable compute-time cost.
\end{abstract}

\maketitle

\section{Introduction}
Simulating interacting quantum many-body systems is one of the key challenges in condensed matter physics, primarily due to the exponential scaling of the Hilbert space with system size. Over the past decades, significant progress has been made through various numerical techniques. Among these, tensor network methods, particularly the density matrix renormalization group (DMRG) \cite{White_1992,Schollwoeck_2011}, have proven especially powerful for one-dimensional systems with low entanglement. These methods have been extended to access thermal states, dynamics, and more complex geometries \cite{Vidal2004, White_2004, Daley_2004,Feiguin_2005, Schmitteckert_2004, Vidal_2007, Kennes_2016}, solidifying their role as a cornerstone of modern quantum many-body simulations in low-dimensional systems.

In disordered systems, especially, robust numerical methods are essential to complement analytical and experimental approaches. Since Anderson’s seminal work on localization in non-interacting systems \cite{Anderson_1958}, significant progress has been made in understanding how disorder suppresses transport. However, when interactions are introduced, the resulting many-body localization (MBL) problem becomes much more intricate \cite{Fleishman_1980,Basko_2006,Gornyi_2005, Znidaric_2008, Oganesyan_2007,Pal_2010, Ros_2015,Kj_ll_2014, Luitz_2015, Luitz_2017, Nandkishore_2015, Altman_2015, Serbyn_2015,Abanin_2019, Morningstar_2022, Laflorencie_2022, Scoquart_2024, Colbois_2024}. Although one-dimensional MBL can be simulated well using conventional DMRG \cite{Lim_2016, Doggen_2018}, the situation in higher dimensions remains less explored \cite{Doggen_2021, Dziarmaga_2022}. Questions surrounding the stability of MBL in 2D at finite energy density and in the thermodynamic limit are still open, and numerical studies have been constrained by the rapid growth of entanglement and the computational cost of simulating large 2D systems.

Traditional tensor network approaches such as DMRG, based on one-dimensional matrix product states (MPS), are not well-suited for capturing the entanglement structure of 2D systems. This limitation has led to an increase in interest in alternative tensor network architectures. Most established are projected entangled pair states (PEPS) that require expensive calculations of the environment \cite{Verstraete_2008,Orus_2019, Cirac_2021}. In the context of disorder, tree tensor networks (TTNs)~\cite{Gunst_2018, Gunst_2018a, Felser_2020, Felser_2021, Felser_2021a, Kloss_2020, Arceci_2022, Jovcheva_2025} offer an alternative compelling approach: Their hierarchical and branching structure naturally accommodates more complex entanglement patterns based on the underlying disorder landscape and enables simulations on larger two-dimensional lattices. Thus, TTNs provide a promising pathway for exploring disordered systems in 2D, where conventional methods struggle to scale.

The paper is set up as follows: In Section~\ref{section: model} we introduce the model. In Section~\ref{section: MPS} we introduce the matrix product state notation to then move on and define tree tensor networks for 2D systems in Section~\ref{section: TTN}. We present a benchmark and results that serve as a proof of concept, including an estimate for the ergodicity-to-localization crossover, in Section~\ref{section: Results} and conclude with a discussion in Section~\ref{section: Discussion}.

\section{The Model}\label{section: model}
We investigate the disordered spin-$\frac{1}{2}$~Heisenberg model. Its Hamiltonian on a two-dimensional square lattice is given by
\begin{align}\label{eq:heisenberg}
    \hat{H} & = \sum_{\langle i,j \rangle} \frac{J}{2}\left( S^+_i S^-_j + S^-_i S^+_j \right) + \Delta S^z_i S^z_j  + \sum_i h_i S^z_i,
\end{align}
where the terms represent nearest-neighbor spin flips ($S_{i}^{+}S_{j}^{-} + \text{h.c.}$) and Ising-like $S^{z}$ interactions on each bond.
$i$ and $j$ label the sites on a two-dimensional square lattice with $\left\langle i,j \right\rangle$ denoting nearest-neighbor pairs. $J$ is the exchange coupling which we set to $J=1$. $\Delta$ is the anisotropy in the $S^{z}S^{z}$ interaction (so $\Delta=1$ corresponds to the isotropic Heisenberg model) and $h_{i}$ is a static random magnetic field at site $i$, drawn uniformly from $[-h, h]$. We will focus on the isotropic case $J=\Delta=1$.
At weak disorder, the system is generally expected to be ergodic, with transport and thermalization governed by the eigenstate thermalization hypothesis (ETH).
At stronger disorder, numerical and experimental studies in one dimension have established the emergence of a many-body localized regime (at least at finite times and sizes), characterized by nonergodic dynamics, area-law entanglement in highly excited states, and Poisson-distributed energy level statistics \cite{Smith_2016, Luitz_2015,Znidaric_2008,Laflorencie_2022}.
In two dimensions, the situation is less settled.
Finite-size simulations and cold-atom experiments have reported signatures reminiscent of MBL, such as slow dynamics, suppressed transport, and deviations from ETH, but the stability of such a phase in the thermodynamic limit remains debated \cite{Doggen_2020, Hur_2025}. Furthermore, some theoretical arguments suggest that thermal “avalanches” can always destabilize localization in the thermodynamic limit \cite{Morningstar_2022} regardless of dimensionality \cite{Hur_2025}.

\section{Matrix Product States}\label{section: MPS}
Matrix product states (MPS) emerged as the mathematical framework underpinning the success of the density-matrix renormalization group (DMRG) algorithm, introduced by White \cite{White_1992, White_1993}. MPS represent quantum states of one-dimensional systems efficiently by leveraging their low entanglement structure.
Consider an arbitrary quantum state for a one-dimensional chain of $N$ sites, written as:
\begin{equation}
    \Ket{\Psi} = \sum_{\sigma_{1} \ldots \sigma_{N}} c_{\sigma_{1}\sigma_{2} \ldots \sigma_{N}}\Ket{\sigma_{1}, \sigma_{2} \ldots \sigma_{N}},
\end{equation}
where $\sigma_{j}$ labels the local basis on site $j$ (with dimension $d$) and $c_{\sigma_{1}\sigma_{2} \ldots \sigma_{N}}$ is the coefficient tensor of size $d^N$. A matrix product state represents this coefficient tensor as a product of tensors:

\begin{equation}
\Ket{\Psi} = \sum_{\substack{\sigma_{1} \ldots \sigma_{N}\\ l_{1} \ldots l_{N-1}}} A_{1,l_{1}}^{\sigma_{1}}A_{l_{1}, l_{2}}^{\sigma_{2}}\ldots A^{\sigma_{N-1}}_{l_{N-2}, l_{N-1}} A^{\sigma_{N}}_{l_{N-1},1}\Ket{\sigma_{1} \ldots \sigma_{N}},
\end{equation}
where $A^{\sigma_{i}}_{l_{i-1},l_{i}}$ are rank-$3$ tensors (except at the boundaries, where they are rank-$2$) with physical index $\sigma_{i}$ and bond indices $l_{i-1}$, $l_{i}$.
In tensor network notation, the coefficient tensor is depicted as
\begin{equation}
    c_{\sigma_{1}\sigma_{2} \ldots \sigma_{N}} = A^{\sigma_{1}}A^{\sigma_{2}}\ldots A^{\sigma_{N}} = \vimg{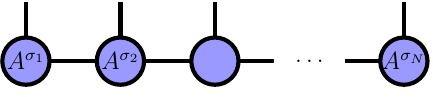},
\end{equation}
which is also the same diagram we use for the full MPS as the basis is fixed:
\begin{equation}
    \Ket{\Psi} = \vimg{mps_1.pdf}.
\end{equation}
The MPS can represent quantum states exactly if the dimensions $\chi_{i}$ of the matrices $A^{\sigma_{i}}$ are not restricted.
However, for a chain of $N$ sites, the bond dimension at the middle of the chain ($\chi_{N/2}$) generally grows exponentially (up to $d^{N/2}$) for highly entangled states to achieve an exact representation.
In practice, the maximum bond dimension $\chi$ is limited to a fixed value, leading to a polynomial scaling of computational cost, typically $\mathcal{O}(N \chi^3)$ for algorithms such as DMRG, compared to the exponential $\mathcal{O}(d^N)$ for the full state. This approximation is effective for states with low entanglement, where the singular values (or Schmidt coefficients) decay rapidly, allowing truncation with minimal loss of accuracy.

\subsection{MPS for two-dimensional systems}
\begin{figure}
\hspace{-35pt}
    \begin{subfigure}[b]{0.2\textwidth}
    \begin{tikzpicture}
        \node[anchor=south west,inner sep=0] (img)
            {\makebox[\linewidth][c]{\vimg[0.3]{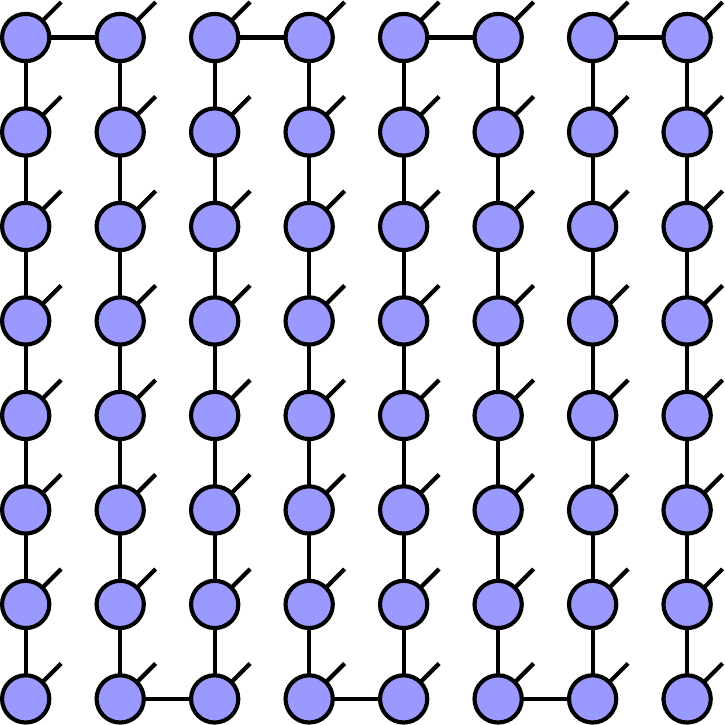}}};
        \node[anchor=north west, xshift=-17pt, yshift=10pt] at (img.north west) {(a)};
    \end{tikzpicture}
    \end{subfigure}
    \hspace{0.03\textwidth}
    \begin{subfigure}[b]{0.2\textwidth}
          \begin{tikzpicture}
        \node[anchor=south west,inner sep=0] (img)
            {\makebox[\linewidth][c]{\vimg[0.3]{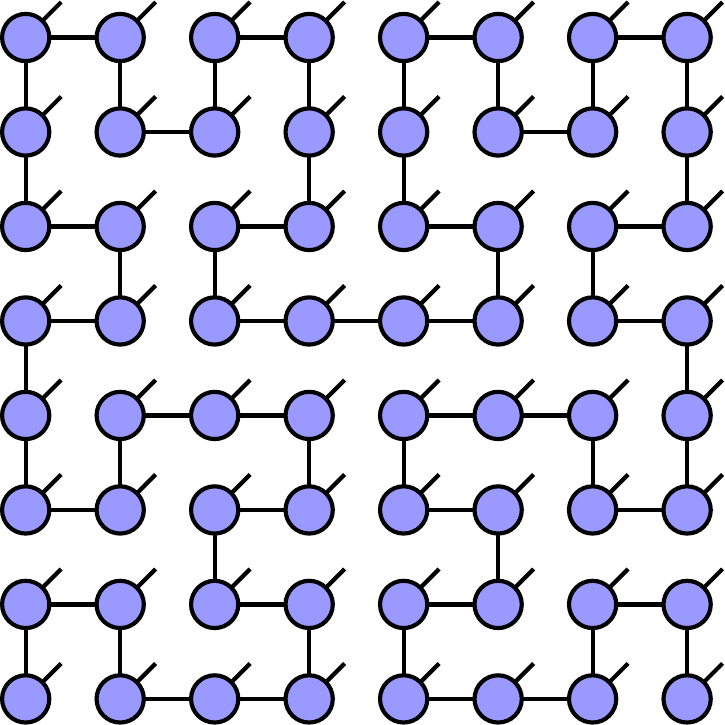}}};
        \node[anchor=north west, xshift=-17pt, yshift=10pt] at (img.north west) {(b)};
    \end{tikzpicture}  
    \end{subfigure}
    
    \caption{MPS mappings for an $L=8$ square lattice. (a) snaking preserves row locality but separates columns by $L$, (b) Hilbert curve maintains $\mathcal{O}(\log L)$ proximity for neighbors.}\label{fig:mps_figures_L8}
\end{figure}

\begin{figure}[htbp!]
    \begin{center}
        \vimg[0.3]{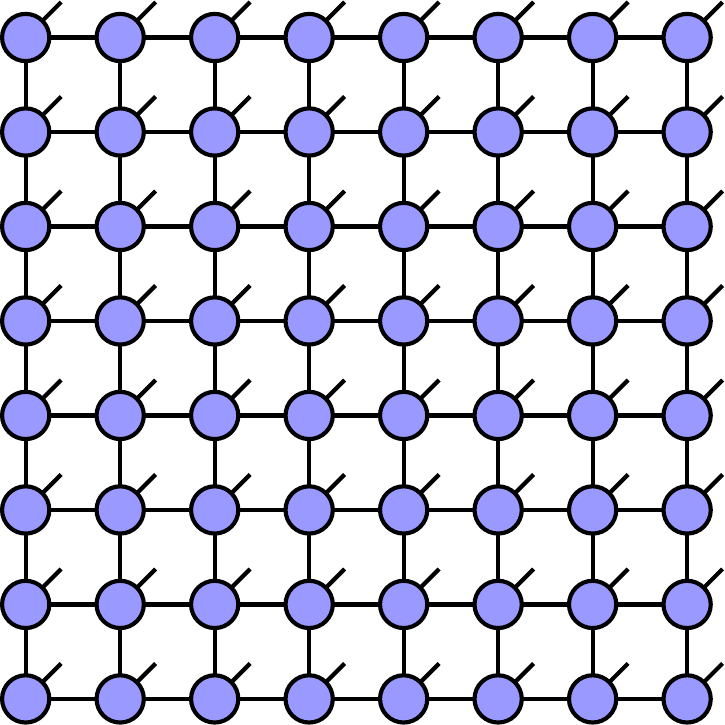}
    \end{center}
    \caption{Projected entangled pair state for a square lattice.}\label{fig:peps_L8}
\end{figure}

Matrix product states (MPS) are natural for one-dimensional chains.
Applying them to a two-dimensional $L\times L$ square lattice is possible, as the derivation of MPS does not require a chain structure.
They therefore only require a reordering of the two-dimensional lattice sites to a chain.
However, this inevitably introduces artificial long-range couplings: nearest neighbors on the lattice can be far apart along the chain.
This affects both the MPO representation of the Hamiltonian (increasing the MPO bond dimension) and the entanglement structure.
A common choice is snake ordering (Fig.~\ref{fig:mps_figures_L8}a)), for which sites are connected in rows and adjacent rows connect at their ends.
However, this  places some vertical neighbors on the original square lattice far apart in the chain.
A mid-chain bipartition corresponds to a vertical cut that crosses $\mathcal{O}(L)$ lattice bonds.
If the states follow an area-law, representing such states with an MPS requires bond dimensions that scale with $\mathcal{O}(\exp(L))$; for volume-law dynamics the situation is even worse.
Likewise, MPOs for nearest-neighbor Hamiltonians acquire couplings whose span grows with $L$, increasing contraction and update costs.
Hilbert space-filling curves (Fig.~\ref{fig:mps_figures_L8}b)) improve the average locality. Most nearest neighbors are mapped closer to each other than for  snaking, and the typical MPO span is reduced.
However, some pairs remain widely separated, and a mid-chain cut still cuts $\mathcal{O}(L)$ bonds.
Thus, Hilbert orderings remain exponentially in scaling.
Projected Entangled Pair States (PEPS) \cite{Verstraete_2008, Orus_2014} (see Fig.~\ref{fig:peps_L8}) encode true 2D connectivity and naturally capture area-law states, but closed loops are introduced that make exact contraction ambiguous and intractable for large system sizes. Practical methods rely on approximate contractions based on environments that become costly for large systems or long real-time evolutions~\cite{Verstraete_2008,Schuch_2007}.

In summary, on the one hand, mapping 2D lattices to 1D chains introduces an unavoidable entanglement bottleneck, while on the other hand, PEPS keep locality but acquire contraction hardness.
This motivates alternative tensor network architectures that balance efficient contraction with improved locality.

\section{Tree Tensor Networks} \label{section: TTN}
Tree tensor networks (TTNs) provide precisely this compromise.
Like MPS, they are variational ansätze built from low-rank tensors, but the tensors are arranged on a loop-free graph (a tree) rather than a chain.
The absence of loops enables exact polynomial-time contractions, and hierarchical structure allows to group nearby sites first, aligning the network with the underlying 2D geometry.
We exploit these features and combine it with a physics-informed structural optimization of the tree layout, to simulate real-time dynamics in disordered two-dimensional systems.

\subsection{Definition and fundamental properties}
\begin{figure}[htbp]
  \centering
  \begin{subfigure}[b]{0.2\textwidth}
      \centering
    \begin{tikzpicture}
        \node[anchor=south west,inner sep=0] (img)
            {\makebox[\linewidth][c]{\vimg[0.3]{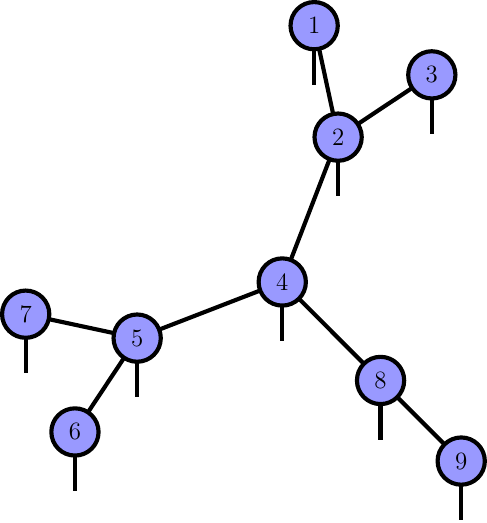}}};
        \node[anchor=north west, yshift=6.5pt] at (img.north west) {(a)};
    \end{tikzpicture}
    \label{fig:basic_tree}
  \end{subfigure}
  \begin{subfigure}[b]{0.2\textwidth}
    \centering
    \begin{tikzpicture}
        \node[anchor=south west,inner sep=0] (img)
            {\makebox[\linewidth][c]{\vimg[0.3]{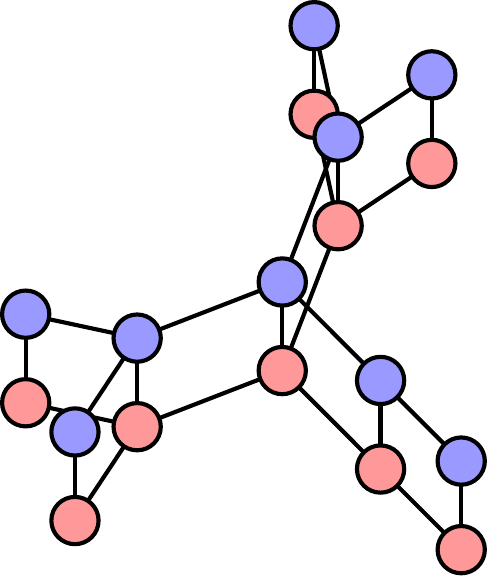}}};
        \node[anchor=north west] at (img.north west) {(b)};
    \end{tikzpicture}
    \label{fig:tree_norm}
  \end{subfigure}

  \vspace{0.5em} 

  \begin{subfigure}[b]{0.2\textwidth}
    \centering
    \begin{tikzpicture}
        \node[anchor=south west,inner sep=0] (img)
            {\makebox[\linewidth][c]{\vimg[0.3]{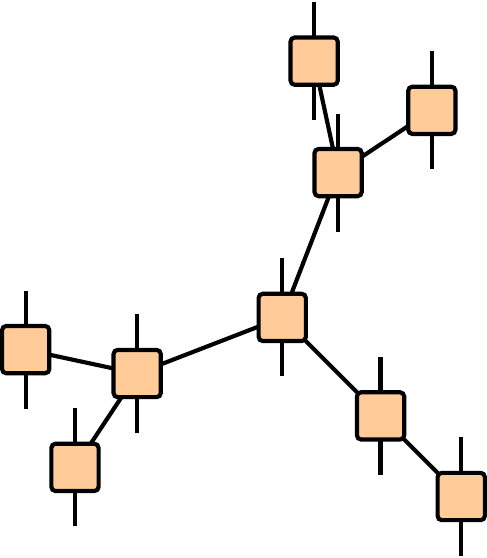}}};
        \node[anchor=north west, yshift=15pt] at (img.north west) {(c)};
    \end{tikzpicture}
    \label{fig:tree_tto}
  \end{subfigure}
  \begin{subfigure}[b]{0.2\textwidth}
    \centering
    \begin{tikzpicture}
        \node[anchor=south west,inner sep=0] (img)
            {\makebox[\linewidth][c]{\vimg[0.3]{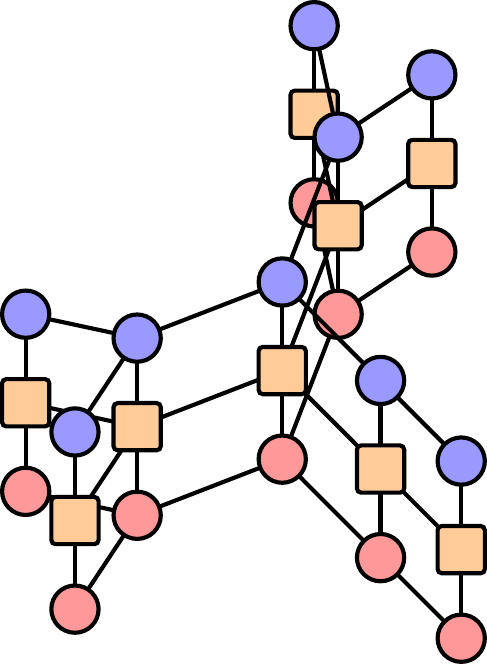}}};
        \node[anchor=north west] at (img.north west) {(d)};
    \end{tikzpicture}
    \label{fig:tree_exp}
  \end{subfigure}

  \caption{(a) Tree tensor network with one physical index per tensor. (b) The norm of a TTN. (c) Extension of the TTN ansatz towards matrix product operators (TTO). (d) The expectation value of an observable.}
  \label{fig:tree_comparison}
\end{figure}
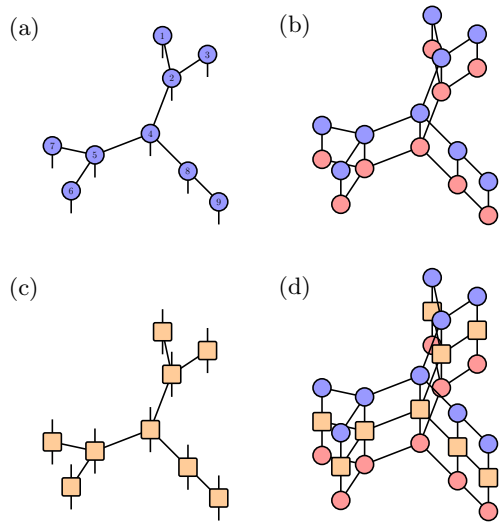

A general tensor network with one physical site for each tensor can be written as
\begin{equation}
    \Ket{\Psi} = \sum_{\substack{\sigma_{1}\ldots \sigma_{N} \\ q_{1} \ldots q_{L}}} A_{Q_{1}}^{\sigma_{1}} A_{Q_{1}}^{\sigma_{2}} \ldots A_{Q_{N}}^{\sigma_{N}} \Ket{\sigma_{1}\ldots \sigma_{N}}
    \label{eq:TTN}
\end{equation}
where $Q_{i} = \left\{q_k: q_{k} \text{ linked to vertex } i \right\}$ is the set of indices connected to node $i$ and $\Ket{\sigma_i}$ are the local states. The tensors $A^{\sigma_{i}}_{Q_{i}}$ can, in principle, have any degree. To build a TTN, we constrain the edges between the vertices to form a tree, i.e. they are connected and have no loops. An example is shown in (Fig. \ref{fig:tree_comparison}a)).
They are particularly suited for systems that have an inherent tree-like structure, but, like MPS, TTNs can also be used for two-dimensional systems. Importantly, an MPS is a special case of a TTN, where the maximum degree of the vertices is $2$. 

Due to the increased number of indices per tensor ($z$ is the coordination number), the size of a TTN also increases as $\mathcal{O}(\chi^{z})$ and with that the algorithmic complexity of exact contraction.
Still, exact contraction remains polynomially scaling in contrast to projected entangled pair states (PEPS), where the exact contraction scales exponentially.

To calculate the norm of a TTN, the external indices of both TTNs are connected just as for the MPS (Fig. \ref{fig:tree_comparison}b)).
The resulting tensor network can then be contracted by starting at the leaf nodes and zipping up the tree to the root nodes in the middle.
This can be achieved with a time complexity of $\mathcal{O}(N \chi^{z+1})$. In analogy to the MPS we can also define tree tensor operators (TTO) for the Hamiltonian (Fig. \ref{fig:tree_comparison}c)).

\begin{figure}[ht]
  \centering
  \begin{overpic}[scale=0.35]{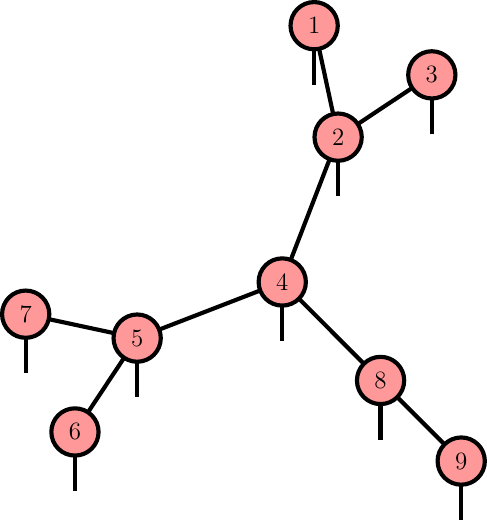}
    \put(2,95){(1)}
  \end{overpic}
  \begin{overpic}[scale=0.35]{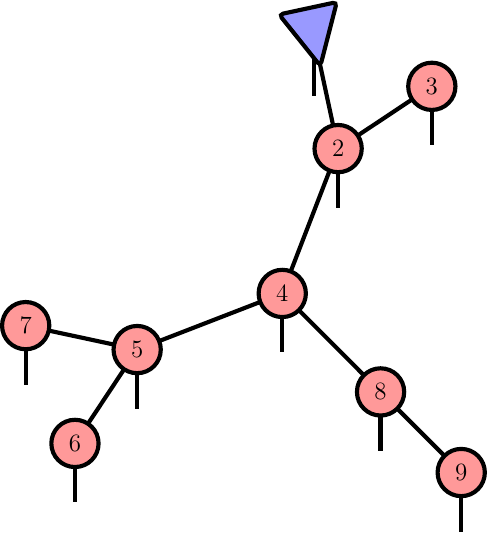}
    \put(2,95){(2)}
  \end{overpic}
  \begin{overpic}[scale=0.35]{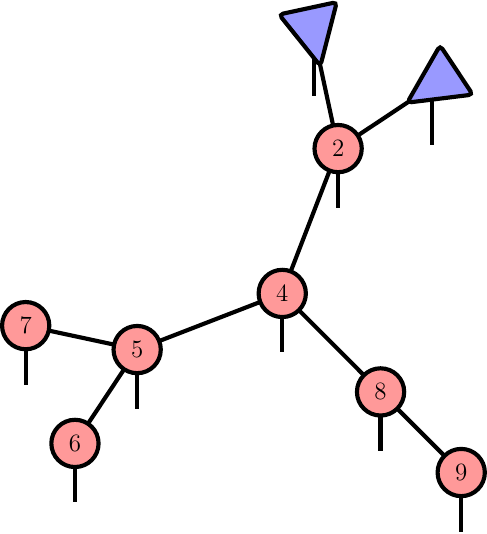}
    \put(2,95){(3)}
  \end{overpic}
  \begin{overpic}[scale=0.35]{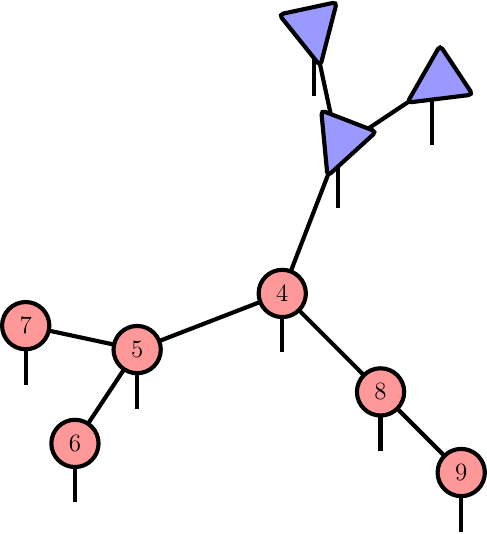}
    \put(2,95){(4)}
  \end{overpic}
  \begin{overpic}[scale=0.35]{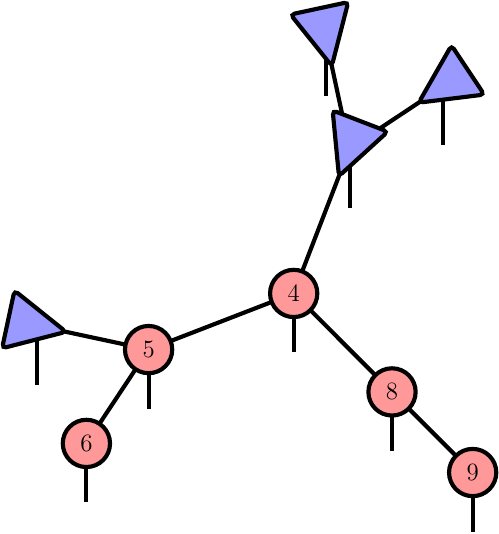}
    \put(2,95){(5)}
  \end{overpic}
  \begin{overpic}[scale=0.35]{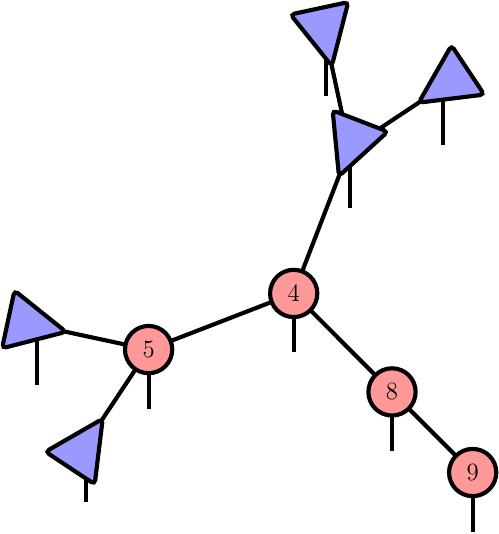}
    \put(2,95){(6)}
  \end{overpic}
  \begin{overpic}[scale=0.35]{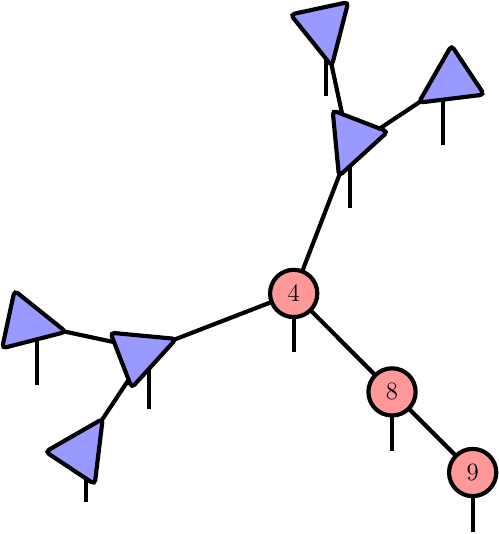}
    \put(2,95){(7)}
  \end{overpic}
  \begin{overpic}[scale=0.35]{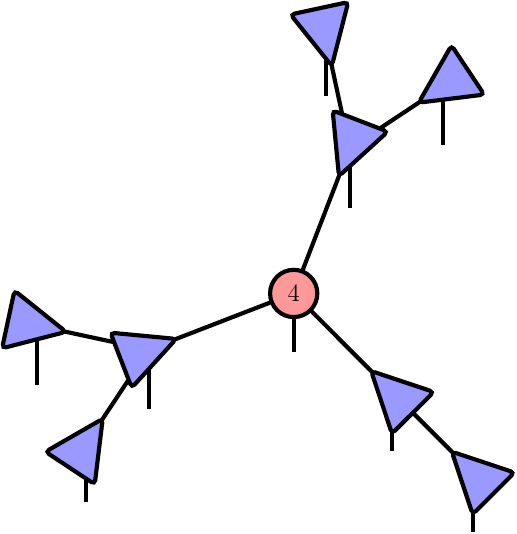}
    \put(2,95){(8)}
  \end{overpic}

  \caption{Step–by–step orthogonalization of the tree tensor network.}
  \label{fig:tree-ortho-sequence}
\end{figure}

\subsection{Orthogonalization and canonical forms}
We can also orthogonalize a TTN with respect to a site or a bond. Starting with our TTN of choice in Fig. \ref{fig:tree-ortho-sequence}, we can orthogonalize it with respect to the middle site by starting at the leaves. We start at an arbitrary leaf node (suppose node $1$) and calculate the SVD (or QR decomposition if we do not need the singular values for, e.g., truncation) of the tensor:
\begin{equation}
    A = U S V^{\dagger}
\end{equation}
$U$ then becomes the new site tensor on the chosen leaf and $S V^{\dagger}$ is contracted with its parent tensor. Before we can continue to orthogonalize site $2$ we first have to orthogonalize all other connected leaves. In this case, the only remaining leave is site $3$. We can then contract $SV^{\dagger}$ from site $3$ with the tensor on node $2$.
Then we can finally orthogonalize node $2$.
To complete the orthogonalization procedure, we can, for example, use the order 6-7-5-9-8 and always orthogonalize in the direction of the root node (for trees there is always only one distinct path).
We then arrive at a TTN with $4$ as its orthogonalization center (see Fig. \ref{fig:tree-ortho-sequence}).

\subsection{Structural flexibility and hierarchical TTNs}

\begin{figure}
    \begin{center}
        \vimg[0.5]{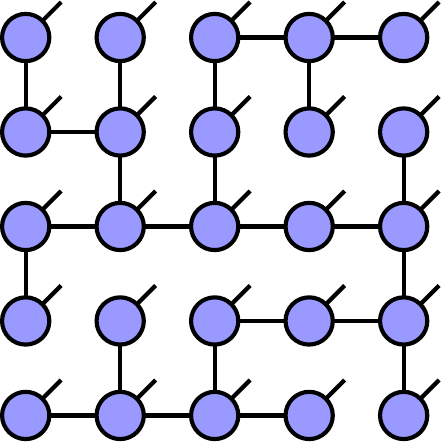}
    \end{center}
    \caption{Example of a TTN with a tensor at each physical lattice site.}\label{fig:tree_lattice}
\end{figure}

There are several ways to map a two-dimensional lattice system to a TTN.
First, we can identify every physical site of the lattice with a tensor and connect them in a way that forms a tree (see Fig.~\ref{fig:tree_lattice}).
This is straightforward but can still lead to long-range interactions between (originally) neighboring lattice sites, where the distance also scales with $\mathcal{O}(L)$.

\begin{figure}[htbp!]
    \centering
    \begin{subfigure}[b]{0.45\textwidth}
        \centering
        \begin{tikzpicture}
            \node[anchor=south west,inner sep=0] (img)
                {\makebox[\linewidth][c]{\includegraphics[height=0.6\textwidth]{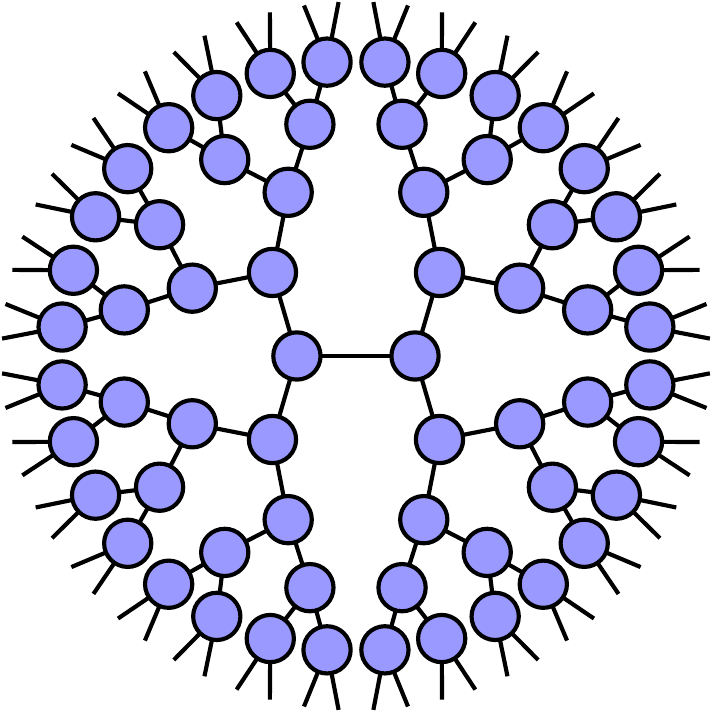}}};
            \node[anchor=north west] at (img.north west) {(a)};
        \end{tikzpicture}
        \phantomsubcaption
        \label{fig:hierarchical_tree_L8_a}
    \end{subfigure}
    \hspace{0.05\textwidth}
    \begin{subfigure}[b]{0.45\textwidth}
        \centering
        \begin{tikzpicture}
            \node[anchor=south west,inner sep=0] (img)
                {\makebox[\linewidth][c]{\includegraphics[height=0.85\textwidth]{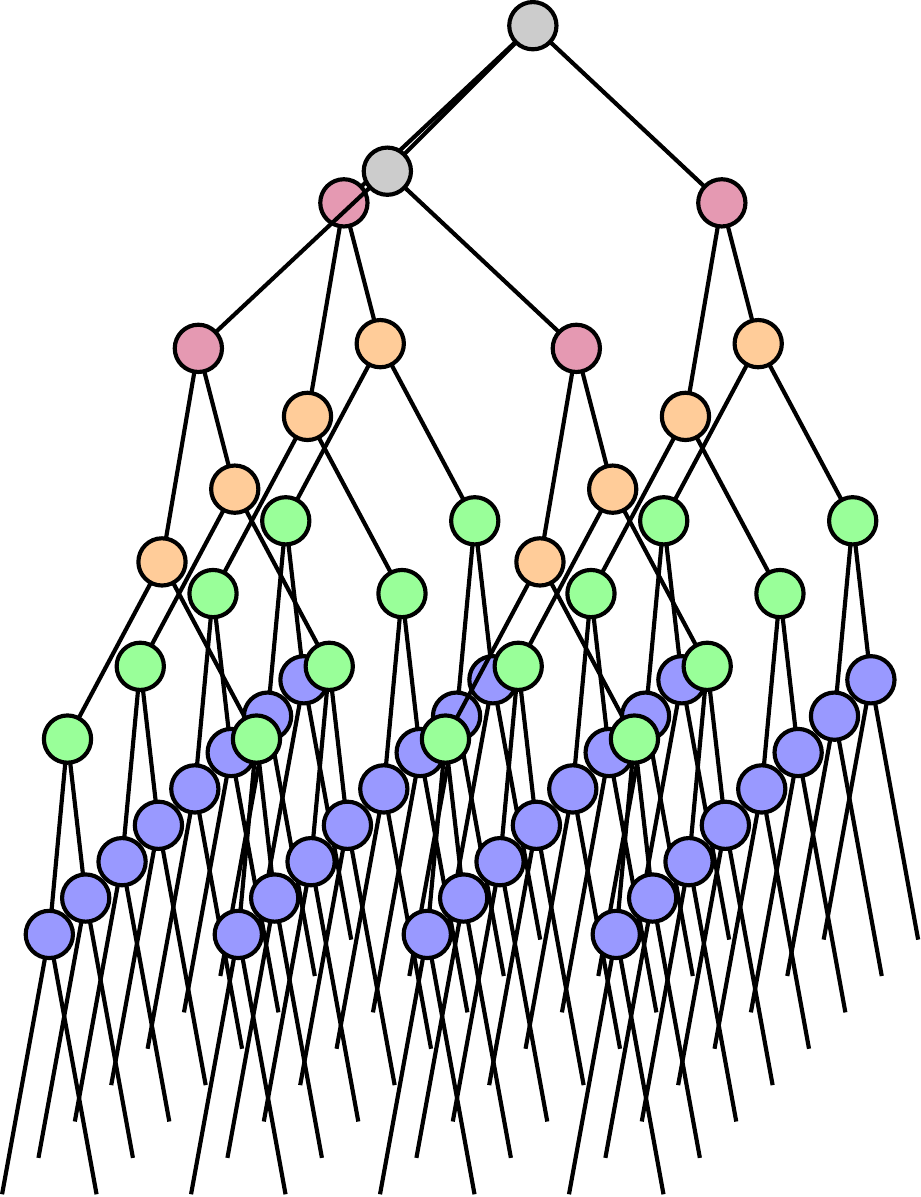}}};
            \node[anchor=north west] at (img.north west) {(b)};
        \end{tikzpicture}
        \phantomsubcaption
        \label{fig:hierarchical_tree_L8_b}
    \end{subfigure}
    \caption{Hierarchical TTN structures for representing a $L=8$ square lattice: (a) a hierarchical tree where the physical indices correspond to open legs at the boundary and (b) the same hierarchical tree arranged to match the geometry of a square lattice.}
    \label{fig:hierarchical_tree_L8}
\end{figure}

On the other hand, we can introduce auxiliary tensors which cannot be attributed to a physical site but serve as connections between other tensors.
One can construct a hierarchical TTN by starting from the physical indices and connecting two of them to one tensor~\cite{Tagliacozzo_2009}.
We then recursively connect two auxiliary tensors to a new tensor in a new layer until all tensors are connected (see Fig.~\ref{fig:hierarchical_tree_L8}).
This way the distance between sites on the tree only scales logarithmically and it provides a better representation of entangled states.

\subsection{Time evolution with TTNs}
We use a variant of the time-dependent variational principle (TDVP) \cite{Haegeman_2011, Haegeman_2016} extended to tree tensor networks \cite{Bauernfeind_2019}. 
Consider a closed quantum system with Hamiltonian $\hat H$ and state $|\psi(t)\rangle$
evolving under the time-dependent Schr\"odinger equation (TDSE)
\begin{equation}\label{eq:tdse}
 i \frac{\mathrm d}{\mathrm d t} |\psi(t)\rangle = \hat H |\psi(t)\rangle.
\end{equation}
Restricting $|\psi\rangle$ to a variational manifold $\mathcal{M}=\{\Ket{\psi[\boldsymbol A]}\}$,
the dynamics must be projected onto $\mathcal M$.
The Dirac--Frenkel variational principle
states that the residual is orthogonal to the tangent space $T_{|\psi\rangle}\mathcal M$:
\begin{equation}\label{eq:diracfrenkel}
 \delta \left\| i\frac{\mathrm d}{\mathrm d t}|\psi[\boldsymbol A]\rangle - \hat H |\psi[\boldsymbol A]\rangle \right\| = 0
 \Longleftrightarrow
 i\frac{\mathrm d}{\mathrm d t}|\psi\rangle
 = P_{T_{|\psi\rangle}\mathcal M}\,\hat H |\psi\rangle,
\end{equation}
where $P_{T_{\Ket{\psi}}\mathcal{M}}$ is the projector onto the tangent space of $\Ket{\psi}$.
This introduces coupled ordinary differential equations (ODE) for the parameters $\boldsymbol A(t)$ whose solutions
define the best approximation to the exact dynamics within the tensor network manifold. The equations are typically set up in terms of local tensors $\phi$ that evolve over time with a local $H_{\mathrm{eff}}$.
For the case of TTNs the tangent space projector is given by
\begin{equation}
    \mathcal{P}_{T |\psi_{[A]}\rangle} = \sum_{i=1}^{N} \mathbb{1}_{i} \otimes \bigotimes_{\left\langle j,i \right\rangle} \hat P_{j}^{i} - \sum_{\left\langle i,j\right\rangle} \hat{P}_{i}^{j} \otimes \hat{P}_{j}^{i} \label{eq: projector}
\end{equation}
where instead of only a projector on the left and right, there is one projector for each neighbor of the tensor $i$. $\hat{P}_{i}^{j}$ projects out all branches of the tree not connected to $i$ in the direction of $j$. For example, the fourth term of the first sum is given by
\begin{equation}
  \mathbb{1}_{4} \otimes \hat P_{2}^{4} \otimes \hat P_{5}^{4} \otimes \hat P_{8}^{4} = \quad \vimg[0.3]{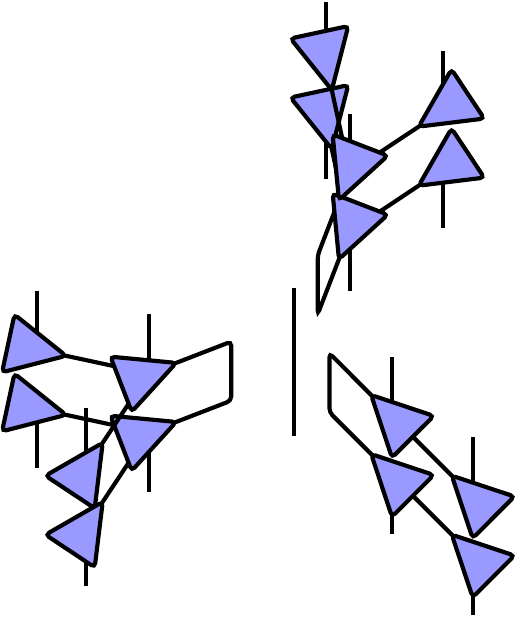} \;,
\end{equation}
where we have an identity on site $4$ and the corresponding projectors to the neighboring nodes in three directions. The first sum in Eq.~\ref{eq: projector} allows arbitrary physical variations of the 
$i$-th tensor, with the neighboring blocks remaining constrained to their canonical form. The terms in the second sum are then responsible for subtracting overcounted gauge transformations on the virtual bonds that were introduced by the first sum.

\subsection{Scaling of the local tensor in two-site TDVP}
\begin{figure}[htbp!]
\begin{equation*}
    \vimg[0.3]{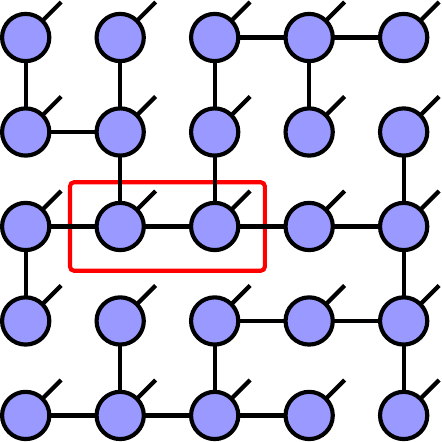} \rightarrow \phi = \vimg[0.3]{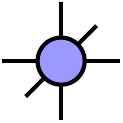} \sim \mathcal{O}_{2\mathrm{s}}(d^{2}\chi^{4}) \;\;\mathrm{vs.}\;\; \mathcal{O}_{1\mathrm{s}}(d \chi^{3})
\end{equation*}

\begin{equation*}
    \vimg[0.3]{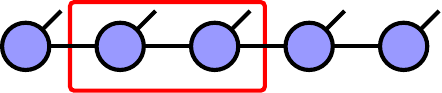}
    \rightarrow
    \phi = \vimg[0.3]{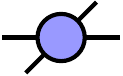}
    \sim \mathcal{O}_{2\mathrm{s}}(d^{2}\chi^{2}) \;\;\mathrm{vs.}\;\; \mathcal{O}_{1\mathrm{s}}(d \chi^{2})
\end{equation*}
\caption{Scaling of the local tensor for the TDVP update in the two-site algorithm ($\mathcal{O}_{2\mathrm{s}}$) in contrast to the one-site algorithm ($\mathcal{O}_{1\mathrm{s}}$).}
\label{fig:complexity-1-2-site}
\end{figure}

A two-site TDVP for TTNs that updates two sites simultaneously and allows for adaptive growth of the bond dimension \cite{Haegeman_2016}, introduces a fundamental difference compared to MPS as the size of the local tensor gets significantly larger. This can be seen in Fig.~\ref{fig:complexity-1-2-site}. Although the size of the tensor only increases by a factor of $d$ when employing two-site instead of single-site TDVP for MPS, for TTNs  there is an additional factor of $\chi$. This greatly increases computational effort, because this is the tensor that also dictates the size of the effective Hamiltonian $H_{\text{eff}}$. The full complexity of $e^{iH_{\text{eff}}} \phi$ depends on the Krylov exponentiation scheme and the contraction order.
A full two-site TDVP for TTNs is therefore computationally expensive, and instead we opt for a hybrid variant in which we start with a two-site TDVP up to a preset bond dimension and then continue with one-site TDVP.
The truncation cutoff of the two-site tensor is kept at a minimum to reduce the error in the early time evolution, as these errors would persist to later time dynamics. With the two-site TDVP the maximum bond dimension is quickly reached at times of the order $t=1$. Increasing the truncation cutoff does not significantly improve on this. Due to this fact, we rely on convergence with respect to the bond dimension to gauge the accuracy of the simulations.

\subsection{Structural optimization}\label{struct_opt}
Hierarchical tree tensor networks offer a topology-agnostic way to simulate a particular system, as they group neighboring indices recursively but do not take into account the underlying disorder.
As the entanglement structure of the systems fundamentally depends on the disorder configuration,
a hierarchical tree is most likely not the ideal structure to model the system, as we would like to connect regions of high entanglement with as few connections as possible.
The number of possible trees spanning a two-dimensional lattice with $n$ physical sites is given by Cayley's formula~\cite{Cayley_1889}.
\begin{equation}
    N  = n^{n-2}\;,
\end{equation}
which makes exploring all trees to find the optimum unfeasible.
Therefore, we have to limit our search for the ideal tree.
First, we restrict our attention to trees with at most three connections to other tensors, since allowing four would raise the algorithmic complexity to the level of most projected entangled pair state (PEPS) methods.

We follow the approach presented in Refs.~\cite{Hikihara_2022, Hikihara_2024, Hikihara_2025}, which we reformulate here for our purpose of representing two-dimensional systems. It uses local updates of the structure:
\begin{itemize}
    \item \textbf{Contract two tensors} of the TTN into a rank-4 tensor $\phi_{ijkl}$:
    \[
        \tikzTreeOpt
    \]
    where the indices can be physical or virtual.

    \item \textbf{Perform several SVDs} of $\phi_{ijkl}$ corresponding to different bipartitions of its indices.
    For example, three possible pairings are:
    \[
        \tikzTreeOptPoss
    \]

    \item \textbf{Compute the entanglement entropy} for each bipartition from the singular values $s_m$:
    \[
        \mathcal{S}_{(ij),(kl)} = - \sum_m s_m^2 \, \log_2 \bigl(s_m^2\bigr),
    \]
    where the singular values are normalized.

    \item \textbf{Select the SVD with the smallest entanglement entropy} and replace the two original tensors by the corresponding decomposition:
    \begin{itemize}
        \item either $U$ and $S V^\dagger$,
        \item or $U S$ and $V^\dagger$,
    \end{itemize}
    depending on the sweeping direction.
\end{itemize}
In this way, we sweep through the TTN and update the bonds sequentially. This minimizes the sum of the bipartite entropies across all bonds
\begin{equation}
    \mathcal{C} (TTN) = \sum_{b \in \text{bonds}} \mathcal{S}_{b}\;,
\end{equation}
with a greedy approach, always choosing the lowest entanglement entropy at every bond. This gives a locally optimal structure.
It is theoretically possible to reach every possible tree structure with these two-site updates, but in practice this is, of course, limited by the entropy structure of the TTN and the structural optimization is likely to get stuck in a local minimum.
Since the entanglement structure is subject to change anyway during the time evolution, this approach nevertheless suffices.
Now we also see why it is advantageous if the number of indices is the same for all the tensors: In this way, we can always group two indices together in the SVD.

\makeatletter
\def\gettval#1{\expandafter\gettval@a#1\relax}
\def\gettval@a#1t=#2.jld2#3\relax{#2}
\makeatother
\newlength{\gutter}\setlength{\gutter}{0.8em}
\newlength{\cell}\setlength{\cell}{(\linewidth - 4\gutter)/5} 
\newcommand{\figscale}{0.7} 

\begin{figure*}[htbp!]
  \centering
  \resizebox{\textwidth}{!}{%
    \begin{minipage}{\textwidth}
      \centering
      \begin{subfigure}{\widthof{\includegraphics[scale=\figscale]{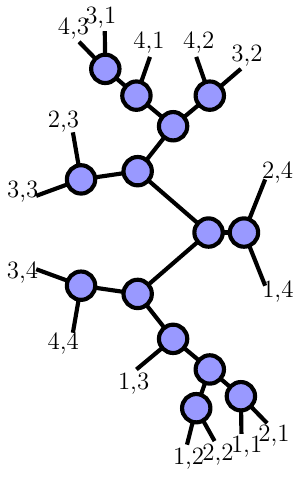}}}
        \includegraphics[scale=\figscale]{tn_tn_t=10.0.jld2.pdf}
        \caption{$t=10$}
      \end{subfigure}\hspace{1em}%
      \begin{subfigure}{\widthof{\includegraphics[scale=\figscale]{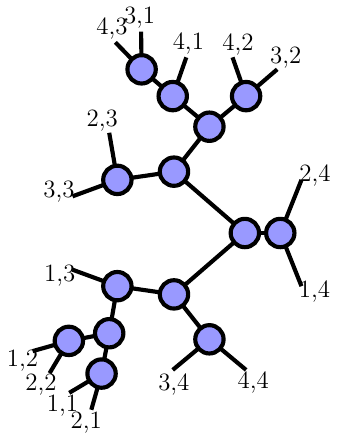}}}
        \includegraphics[scale=\figscale]{tn_tn_t=20.0.jld2.pdf}
        \caption{$t=20$}
      \end{subfigure}\hspace{1em}%
      \begin{subfigure}{\widthof{\includegraphics[scale=\figscale]{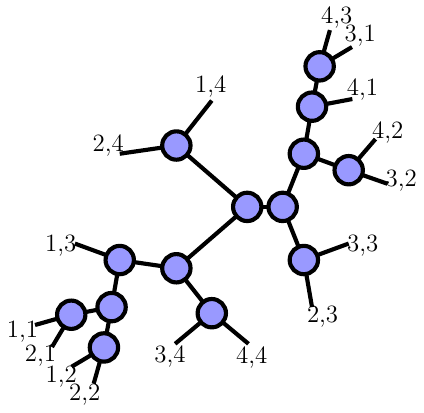}}}
        \includegraphics[scale=\figscale]{tn_tn_t=30.0.jld2.pdf}
        \caption{$t=30$}
      \end{subfigure}\hspace{1em}%
      \begin{subfigure}{\widthof{\includegraphics[scale=\figscale]{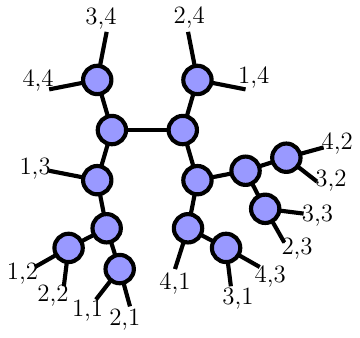}}}
        \includegraphics[scale=\figscale]{tn_tn_t=40.0.jld2.pdf}
        \caption{$t=40$}
      \end{subfigure}%
    \end{minipage}
  }
  \caption{Tensor-network snapshots at different times $t$ for a $4\times4$ system with $h=10$.}
  \label{fig:tn-series}
\end{figure*}

To provide an example in Fig.~\ref{fig:tn-series} the evolution of the optimal tree structure over time is shown for a $4 \times 4$ system. The structure evolves only gradually with changes occurring predominantly in localized regions of the graph. 

\section{Results}\label{section: Results}
\subsection{Benchmark}
\begin{figure*}
\includegraphics[width=\linewidth]{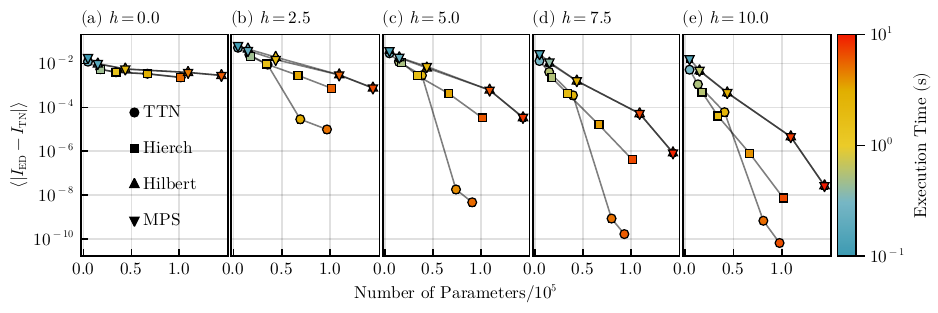}
\caption{
Averaged absolute difference to exact diagonalization results, plotted against the number of parameters. From left to right: Increasing disorder strength (each averaged over $50$ configurations). ``TTN'' denotes a tree tensor network with running structural optimization. ``Hierch'' denotes a hierarchical TTN without structural optimization. ``Hilbert'' denotes a run with an MPS set up like a Hilbert curve, while ``MPS'' denotes a snaking MPS. Performance measurements were conducted on a Intel Xeon 8468 Sapphire CPU and multi-threaded over 4 cores.
}
\label{fig:L4_params_error}
\end{figure*}
We benchmark accuracy against exact diagonalization (ED) on small lattices. Due to the exponential growth of the Hilbert space, accessible system sizes are limited by memory.
The implementation constructs the Hamiltonian in the $S^{z}$ basis within a fixed magnetization sector. For time evolution, we employ Krylov subspace methods to apply the exponential of the Hamiltonian \cite{Hochbruck_1997}. 
We time-evolve an initial product state of the form of a columnar Néel state:
\begin{equation}\label{eq:col_neel}
\Ket{\psi} = \vimg[0.3]{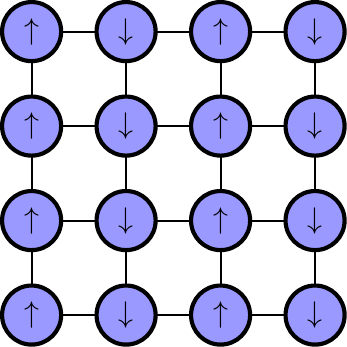}
\end{equation}
and track the columnar imbalance:
\begin{equation}\label{eq:imbalance}
    I(t) = \frac{2}{L^2} \sum_{i=1}^{L} \sum_{j=1}^{L} (-1)^i \,\langle S^z_{\left(i,j\right)}(t)\rangle
\end{equation}
The factor of $2$ normalizes the observable so that the columnar Néel state in Eq.~\eqref{eq:col_neel} has $\mathcal{I}=1$. This quantity serves as a measure of how much memory of the initial state is retained in the system. In the ergodic regime $\mathcal{I}(t)$ quickly approaches $0$, whereas in the MBL regime it remains at a non-zero value.\\
There is one last caveat in assessing the expressiveness of the TTNs. Since TTNs have a higher rank than MPS, i.e.,~a higher number of legs connected to the local tensors, the bond dimension is distributed across more legs. Thus, for a given bond dimension, a local tensor in a TTN carries much more information than the corresponding one in an MPS and, accordingly, is computationally more expensive to contract.
Therefore, instead of comparing the accuracy for a given bond dimension between TTN and MPS, we need to fix the total number of parameters
\begin{equation}
N_{\text{par}} = \sum_{i} \mathrm{size}(A^{i}),
\end{equation}
where the size is given by the product of the dimensions of every index (tensors are dense, no symmetries are exploited). Using this parameter count provides an ansatz-agnostic metric that is more suitable than bond dimension alone. We consider the real-time evolution of the columnar Néel state up to $t=100$ with varying disorder strength of a $4 \times 4$ system.
The execution time is represented by the color scale.
For each data point, we performed simulations for $10$ disorder configurations and averaged the error.
\begin{figure*}
        \includegraphics[width=\linewidth]{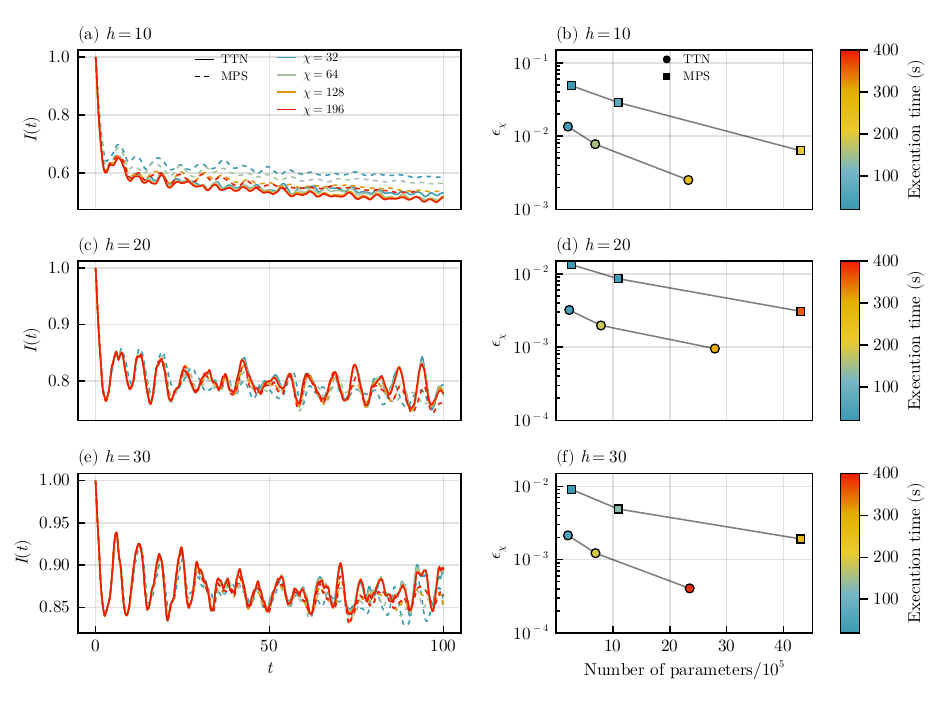}
    \caption{Imbalance over time for a $12 \times 12$ system for different disorder strengths. Left column shows the imbalance for both MPS and TTN simulations for different maximal bond dimensions and for disorder values below and close to the localization crossover. Right column shows convergence with respect to the run with highest bond dimension $\chi_{\mathrm{max}} = 196$. In these reference runs, the TTNs have approximately $60 \cdot 10^5$ parameters in total while the MPS have $90 \cdot 10^5$ parameters.}
    \label{fig:imbalance_12}
\end{figure*}
As shown in Fig.~\ref{fig:L4_params_error}, for a given number of parameters, TTNs show lower errors across disorder strengths. TTN simulations with structural optimisation (labelled ``TTN'' in Fig.~\ref{fig:L4_params_error}) perform particularly well. Hierarchical trees without structural optimization still improve upon the MPS, but are less accurate than the optimized TTNs. The Hilbert curve and the snaking curve as two variants of MPS mappings perform nearly identically.
In the absence of disorder, the imbalance rapidly decays to zero. Larger bond dimensions improve accuracy at short times, but errors converge to similar values at long times. For reference, the maximum bond dimension for an MPS with $N=16$ sites is $2^{N/2}=256$.
This is expected for high entanglement (volume-law scaling) states in the ergodic regime, where substantially larger bonds would be required. Consequently, the error remains approximately constant and the TTNs cannot exploit any physical structure of the system.\\
Overall, the TTN increase the accuracy in the simulation of the disordered system or, equivalently, require fewer parameters for a given target error.
For higher disorder, the accuracy saturates, consistent with entering the localized regime, where increasing disorder no longer reduces entanglement.
At a given accuracy, the run-time is generally better for the TTNs as provided by the color code in Fig.~\ref{fig:L4_params_error}. The increased complexity of the tree tensor networks is fully compensated by the increase in expressive prowess.

\subsection{Imbalance dynamics of larger systems}
The imbalance dynamics in a $12\times 12$ system for a single disorder realization at $h=10, 20, 30$ is shown in Fig.~\ref{fig:imbalance_12}(a),(c),(e).
The system does not show localization for these disorder values, but TTNs are still capable of finding a good representation of the time-evolved state and converge significantly faster than MPS. Convergence is measured as the average absolute difference of the imbalance to the highest bond dimension $\chi_{\mathrm{max}} = 196$, 
\begin{equation}
\epsilon_\chi = \left<\left| I_{\chi_\mathrm{max}} - I_\chi \right|\right>
\end{equation}
The convergence is shown in Fig.~\ref{fig:imbalance_12}(b),(d),(f) as a function of the number of parameters in the tensor network. The number of parameters for the reference run with highest bond dimension differ slightly, but as the convergence is a relative measure, better convergence is interpreted as being closer to the exact solution, since an increase in bond dimensions should not matter in the numerically exact limit. For all bond dimensions and disorder values, TTNs outperform the MPS ansatz. To reach a given convergence, the TTNs require up to a factor of $10$ less parameters already below the localization crossover. The TTN simulations also require less compute time, at least for moderate bond dimension. Increasing the bond dimension can further reduce the error for TTNs, although at additional cost. This means that the complexity of TTNs remains favorable for large-scale disordered 2D systems, needing fewer parameters than MPS for a faithful representation. The physics-informed ansatz further proves especially powerful with increased disorder, as expected.
\subsection{Ergodicity-to-localization crossover}
\begin{figure}
    \centering
    \includegraphics[width=\linewidth]{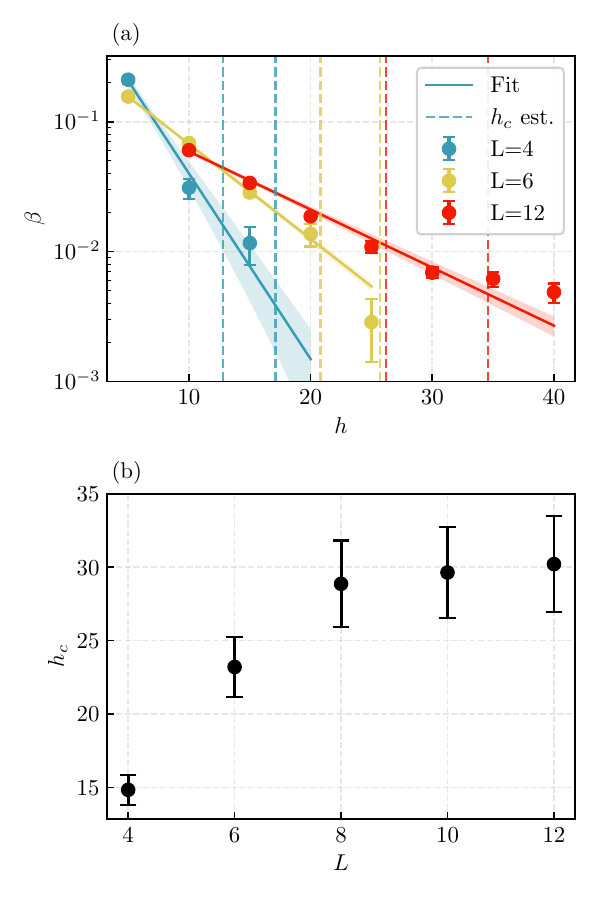}
    \caption{Ergodicity-to-localization crossover as obtained from the imbalance dynamics. (a) Imbalance decay rate $\beta$ as a function of disorder strength $h$ for different system sizes, averaged over $N=200$ disorder realizations, obtained with the bootstrapping procedure described in the main text. Solid lines mark the exponential fit to extract $h_c$ and the vertical dashed lines mark the obtained $h_{1,2}$ that determine $h_c$. (b)~Crossover $h_c(L)$ as a function of system size. Simulations were performed using the TTN ansatz with maximum bond dimension $\chi = 128$.}
    \label{fig:beta_combined}
\end{figure}
We can use the decay of the imbalance as a diagnostic for localization~\cite{Doggen_2020}. In the ergodic regime, $I(t)$ typically decays as a power law,
\begin{equation}
    I(t) \sim t^{-\beta},
\end{equation}
with exponent $\beta > 0$. In contrast, in the localized regime, the imbalance saturates at a non-zero value at long times, corresponding to $\beta \to 0$.
The fitting procedure is performed in accord to Ref.~\cite{Doggen_2020}.
We extract the power-law decay exponent $\beta$ from the disorder-averaged imbalance $I(t)$ by fitting the ansatz $I(t) \propto t^{-\beta}$ within the time window $t \in [30, 100]$. To estimate the uncertainty $\sigma_\beta$, we employ a bootstrap approach based on resampling the individual disorder realizations. We generate synthetic datasets by sampling $N=200$ disorder realizations with replacement, and,
for each bootstrap sample, we compute the new disorder-averaged mean imbalance and perform the power-law fit.
The exponent $\beta$ and its error are determined from the mean and standard deviation of the resulting distribution of fitted exponents.
To estimate the critical disorder strength $h_c(L)$ for a given system size $L$ (note, that this can only be considered a finite-size crossover scale), we analyze the dependence of the decay exponent $\beta$ on the disorder field $h$.
This is modeled by an exponential decay $\beta(h) = A \exp(B h)$.
As this only converges to zero instead of crossing it, we solve for $\beta(h_{1,2}) = \delta\beta_{1,2}$ with $\delta\beta_{1,2} = 0.01, 0.005$. The critical disorder is given by the mean, $h_c=(h_1 + h_2)/2$, and the error estimate corresponds to $h_1$ and $h_2$, respectively. 

Representative results are shown in Fig.~\ref{fig:beta_combined}(a).
For the smallest system size ($L=4$), $\beta$ decreases rapidly and approaches zero at moderate disorder, consistent with a crossover to localized dynamics. For larger $L$, the decrease of $\beta$ with $h$ is slower, indicating that increased disorder strengths are necessary to localize the system. We estimate the critical disorder $h_c$ from the $h_{1,2}$ given by the vertical dashed lines in Fig.~\ref{fig:beta_combined}(a)). Fig.~\ref{fig:beta_combined}(b) shows the critical disorder as a function of system size. For the system sizes shown, there is a slowdown but no clear convergence to a finite $h_c$.
This behavior is in line with the avalanche scenario, which predicts that true MBL in two dimensions is unstable to thermal inclusions and would only be realized at infinite disorder \cite{Morningstar_2022, Hur_2025}. Our TTN-based results are therefore best interpreted as evidence of a finite-size crossover into localization-like, slow dynamics and further support the literature findings that the crossover $h_c(L)$ shows a slow (possibly logarithmic) divergence with $L$ \cite{Doggen_2020, Decker_2022, Herre_2023, Scoquart_2024}.

\section{Discussion}\label{section: Discussion}
We introduced a physics-informed tensor network based ansatz to simulate the many-body-localization problem in two dimensions. Instead of constructing a simple matrix product state with long-range interactions or using PEPS, we set up a tree tensor network that is structurally optimized to follow high- and low-entanglement regimes in the disordered system. This proved to be an efficient way to compress the full many-body wave function based on entanglement. Our results show how this can extend the reach of real-time simulations both in system size and maximal simulation time, even at moderate disorder strength. The additional complexity overhead of the TTNs over MPS turns out to be overcompensated by the increased expressiveness of the ansatz.
We then used the ansatz to perform an analysis of the ergodicity-to-localization crossover, and found our results to be consistent with previous statements about the absence of a localized phase in the thermodynamic limit in two dimensions.\\
The findings show that TTNs capture two-dimensional entanglement patterns more effectively than MPS.
Runtime scales roughly with the total number of variational parameters; the additional overhead from the tree contractions remains small and one can reach significantly higher accuracy at similar runtimes compared to MPS simulations.
In addition, entanglement-guided structural optimization reduces bipartite entropies across bonds, aligning the tree with the evolving entanglement and improving stability during time evolution.\\
A natural next step is to exploit symmetries directly in the tensor network. For the disordered Heisenberg model with fields along $z$, enforcing the global $U(1)$-symmetry via charge-labeled, block-sparse tensors would confine the dynamics to the correct sector, reduce effective parameter counts, and yield block-diagonal effective Hamiltonians for TDVP updates. Symmetry-preserving truncations and symmetry-aware bond expansions should further optimize bond growth. GPU offloading dominant linear-algebra kernels is another plausible way to reduce compute-time.
The landscape of many-body localization in two dimensions remains challenging. Even with improved numerics, distinguishing finite-time, finite-size slow dynamics from genuine asymptotic localization is subtle, and rare-region effects complicate interpretation. The methods developed here, TTN+TDVP with structural optimization, provide a much needed compromise between MPS and PEPS to probe larger systems and longer times, but definitive statements about the stability of 2D MBL will require pushing scales much further. Altogether, these results position TTN-based TDVP as a scalable tool for real-time studies of disordered two-dimensional quantum systems and outline a clear path for pushing simulations to regimes that were previously out of reach.

\begin{acknowledgements}
D.M.K. and J.-N.H. acknowledge funding by the Deutsche Forschungsgemeinschaft (DFG, German
Research Foundation) – 508440990. Computations were performed with computing resources granted by RWTH Aachen University under project thes2506.
\end{acknowledgements}
\section*{Data availability}
The data supporting the findings of this article are not
publicly available. The data are available from the authors
upon reasonable request.

\FloatBarrier

\end{document}